\documentclass[12pt]{article}
\usepackage{epsf}
\usepackage{psfig}
\usepackage{epsfig}
\usepackage{graphicx}
\usepackage{amssymb}

\def\di{\displaystyle}

\def\eq#1{(\ref{#1})}
\def\Eq#1{Eq.~(\ref{#1})}

\def\tr{{\rm tr}}

\def\s0#1#2{\mbox{\small{$ \frac{#1}{#2} $}}}
\def\0#1#2{\frac{#1}{#2}}

\def\eq#1{(\ref{#1})}
\def\Eq#1{Eq.~(\ref{#1})}

\def\di{\displaystyle}

\def\bg{\begin{eqnarray}\begin{array}{rcl}\displaystyle}
\def\eg{\end{array} &\di    &\di   \end{eqnarray}}
\def\bm#1{\begin{eqnarray}\begin{array}{#1}\di}
\def\bmo#1{\begin{eqnarray*}\begin{array}{#1}\di}
\def\bml#1#2{\begin{eqnarray}\begin{array}{#1}\label{#2}\di}
\def\bgo{\begin{eqnarray*}\begin{array}{rcl}\displaystyle}
\def\ego{\end{array} &\di    &\di \nonumber  \end{eqnarray*}}

\def\btensor#1#2{\renew\left#1\begin{array}{#2}\di}
\def\brtensor#1#2#3{\ren#3\left#1\begin{array}{#2}}
\def\botensor#1#2{\renew\left#1\begin{array}{#2}}
\def\etensor#1{\end{array}\right#1}

\def\eq#1{(\ref{#1})}
\def\Eq#1{Eq.~(\ref{#1})}

\def\tr{{\rm tr}}

\def\T{{\rm T}}

\def\be{\begin{equation}}
\def\ee{\end{equation}}
\def\bea{\begin{eqnarray}}
\def\eea{\end{eqnarray}}

\def\h{\hbox{$\frac{1}{2}$}}

\font\af=msbm12
\def\T{{\mbox{\af T}}}
\def\R{{\mbox{\af R}}}

\date{\today}

\def\ren#1{\renewcommand{\arraystretch}{#1}}

\def\renew{\renewcommand{\arraystretch}{1}}
\ren{1.5}

\begin{document}


\begin{flushright}
INLO-PUB-03/02\\
FAU-TP3-02-18\\
hep-th/0205116
\end{flushright}
\par
\vskip .5 truecm
\large \centerline{\bf Constituents of Doubly Periodic Instantons}
\par
\vskip 0.5 truecm
\normalsize
\begin{center}
{\bf C.~Ford} ${}^{a}$ and
{\bf  J.~M.~Pawlowski} ${}^{b}$
\\
\vskip 0.5 truecm
${}^a$\it{
Instituut-Lorentz for Theoretical Physics\\
Niels Bohrweg 2, 2300 RA Leiden, The Netherlands\\ 
{\small\sf ford@lorentz.leidenuniv.nl}\\
}
                                            
\vskip 0.5 truecm
${}^b$\it{Institut f\"ur Theoretische Physik III,
Universit\"at Erlangen\\
Staudtstra\ss e 7,
D-91058 Erlangen, Germany \\
{\small\sf jmp@theorie3.physik.uni-erlangen.de}
}

\vskip 0.5 true cm

\end{center}

\vskip .7 truecm\normalsize
\begin{abstract} 

Using the Nahm transform we investigate 
doubly periodic charge one $SU(2)$ instantons with radial symmetry.
Two special points 
where the Nahm zero modes have softer singularities
are identified as constituent locations.
To support this picture, the action density is computed
analytically and numerically within a two dimensional slice
containing the two constituents.
For particular values of the parameters
the torus can be cut in half yielding two copies of a twisted
charge $\frac{1}{2}$ instanton. Such objects comprise a single
constituent.

\end{abstract}
\baselineskip=20pt Topologically non-trivial objects play a pivotal
r${\rm \hat o}$le in most confinement scenarios. A prominent example
being the dual Meissner effect via the condensation of magnetic
mono\-poles.  To date, there is no gauge independent way of
identifying the monopole content of a given gauge potential.  However,
some progress has been made with regard to instantons.  In particular,
charge one $SU(N)$ calorons, or periodic instantons, can be viewed as
bound states of $N$ monopole constituents
\cite{Lee:1998bb,Kraan:1998pm}.  This identification does not hinge on
a particular gauge choice as the constituents are clearly visible as
peaks in the action density.  Presumably, similar results hold for
`higher' tori, that is, instantons on $\T^2\times\R^2$ (doubly
periodic instantons), $\T^3\times\R$ and the four torus $\T^4$.  In
the latter case twists must play a r${\rm \hat o}$le since untwisted
charge one instantons do not exist \cite{Braam:1988qk}.

There is an almost complete lack of explicit results for multiply
periodic instantons in the literature.  However, the \it existence \rm
of higher charge instantons on $\T^4$ was established by Taubes.  More
recently, doubly periodic charge one instantons have been discussed
\cite{Jardim:1999dx,Ford:2000zt}. In \cite{Ford:2000zt} doubly
periodic charge one $SU(2)$ instantons with radial symmetry in the
non-compact $\R^2$ directions were considered.  Under the Nahm
transformation these instantons are mapped to abelian potentials on
the dual torus $\tilde\T^2$.  Our basic approach is to start with
these rather simple abelian potentials and then Nahm transform to
recover the original $SU(2)$ instanton. This involves solving certain
Weyl-Dirac equations; for each $x$ in $\T^2\times\R^2$ one has a
different Weyl equation on $\tilde\T^2$.  The Weyl zero modes were
determined explicitly for a two dimensional subspace
\cite{Ford:2000zt} of $\T^2\times\R^2$ (this subspace corresponds to
the origin of $\R^2$), see also \cite{Reinhardt:2002cm}.

Although this falls short of a complete solution, this subspace is, by
virtue of the radial symmetry, exactly where any constituents are
expected to lie.  In this letter we use the explicit zero modes to
develop a constituent picture of the associated doubly periodic
instantons.  There are two points in the subspace where the zero modes
have different singularity profiles. These are obvious candidates for
constituent locations.  The behaviour of the field strengths in the
vicinity of the proposed constituent locations is investigated
analytically and numerically.  It is possible to arrange that the two
constituents are identical lumps. If we further fix their separation
to be a half period, the torus can be cut in two.  In each half of the
torus we have a twisted instanton of topological charge $\frac{1}{2}$
comprising a single constituent.

Before we specialise to $\T^2\times\R^2$ let us briefly recall how the
Nahm transform is formulated on $\T^4$ \cite{Braam:1988qk}.  Start
with a self-dual anti-hermitian $SU(N)$ potential, $A_\mu(x)$, on a
euclidean four-torus with topological charge $k$.  A gauge field on
$\T^4$ is understood to be an $\R^4$ potential which is periodic
(modulo gauge transformations) with respect to $x_\mu\rightarrow
x_\mu+L_\mu \quad (\mu=0,1,2,3)$, the $L_\mu$ being the four periods
of the torus.  The next, apparently trivial, step is to turn the
$SU(N)$ instanton into a $U(N)$ instanton by adding a constant $U(1)$
potential, $A_\mu(x)\rightarrow A_\mu(x)-iz_\mu$, where the $z_\mu$
are real numbers.  We can regard the $z_\mu$ as coordinates of the \it
dual torus \rm, $\tilde\T^4$, since the shifts $z_\mu\rightarrow
z_\mu+2\pi/L_\mu$ can be effected via \it periodic \rm $U(1)$ gauge
transformations.

Now consider the $U(N)$ Weyl operator
\begin{equation}
D_z(A)=\sigma_\mu D^\mu_z(A),~~~~
D_z^\mu(A)=\partial^\mu+A^\mu(x)-iz^\mu,
\end{equation}
$\sigma_\mu=(1,i\tau_1,i\tau_2,i\tau_3)$ where the $\tau_i$ are Pauli
matrices.  Provided certain mathematical technicalities are met
$D^\dagger_z(A)=-\sigma_\mu^\dagger D_z^\mu(A)$ has $k$
square-integrable zero modes $\psi^i(x;z)$ with $i=1,2,...,k$.  The
Nahm potential is defined as
\begin{equation}\label{nahm}
\hat A_\mu^{ij}(z)=\int_{T^4}d^4x\,
{\psi^i}^\dagger(x;z)\frac{\partial}{\partial z^\mu}
\psi^j(x;z).\label{basicnahm}
\end{equation}
Here the zero modes are taken to be orthonormal.
Remarkably,
$\hat A(z)$ is a $U(k)$ instanton on the dual torus with topological charge
$N$. If we execute a second Nahm transformation on
$\hat A$, the original $SU(N)$ instanton will be recovered.

Formally, one can obtain the $\T^2\times\R^2$ Nahm transform by taking
 two of the periods, say
$L_0$ and $L_3$, to be infinite.
Given an $SU(N)$ instanton periodic with respect to
$x_1\rightarrow x_1+L_1$, $x_2\rightarrow x_2+L_2$ its Nahm
transform is
\begin{eqnarray}
\hat A_\mu^{ij}(z)&=&\int_{T^2\times R^2} d^4x\,
\psi^i{}^\dagger(x;z)\frac{\partial}{\partial z^\mu}\psi^j(x;z)
~~~~~~~\mu=1,2\label{t2nahm}\nonumber
\\
\hat A_\mu^{ij}(z)&=&\int_{T^2\times R^2} d^4x\,
\psi^i{}^\dagger(x;z)ix_\mu \psi^j(x;z)
~~~~~~~\mu=0,3
\end{eqnarray}
where the $\psi^i(x;z)$ with $i=1,...,k$
are orthonormal zero modes of 
$D_z^\dagger(A)$.
Note that we may gauge $z_0$ and $z_3$ to zero.
The Nahm potential is a self-dual 
$U(k)$ gauge field 
 with $N$ singularities in $\tilde\T^2$ (periods
$2\pi/L_1$ and $2\pi/L_2$).

One can regard $\hat A_1(z)$ and $\hat A_2(z)$
as the components of a two dimensional
gauge potential, and combine
$\hat A_0(z)$ and $\hat A_3(z)$ into a `Higgs' field
$\Phi(z)=\frac{1}{2}(\hat A_0(z)-i\hat A_3(z))$.
The next step is to seek solutions of the dimensionally reduced
self-duality (or Hitchin) equations.
In the one-instanton sector this is rather straightforward since the
 corresponding Nahm potential is abelian.
What is more tricky, however,
is to execute the second Nahm transformation to recover the instanton
itself.
We shall
 restrict ourselves  to the special case of a zero Higgs field.
This means that the corresponding instanton will
be radially symmetric: 
local gauge invariants such as the action density
 depend on $x_1$, $x_2$ and $r=\sqrt{x_0^2+x_3^2}$
only.
With this restriction the self-duality equations
are just
\begin{equation}
\hat F_{y\bar y}=\partial_y \hat A_{\bar y}-\partial_{\bar y}\hat A_y=0.
\label{F=0}\end{equation}
Here we have used complex coordinates;
$y=z_1+iz_2$, $\bar y=z_1-iz_2$,
$\partial_y=\frac{1}{2}(\partial_{z_1}-i\partial_{z_2})$,
$\partial_{\bar y}=\frac{1}{2}(\partial_{z_1}+i\partial_{z_2})$,
$\hat A_y=\frac{1}{2}(\hat A_1(z)-i\hat A_2(z))$
and
$\hat A_{\bar y}=\frac{1}{2}(\hat A_1(z)+i\hat A_2(z))$. Consider the ansatz
\begin{equation}
\hat A_y=\partial_y\phi,~~~~~~~~
\hat A_{\bar y}=-\partial_{\bar y}\phi,
\end{equation}
which gives
$\hat F_{y\bar y}=-2\partial_y\partial_{\bar y}\phi$. Then, in order to 
satisfy \eq{F=0}, 
$\phi$ must be harmonic except at two singularities, 
since we are aiming for an $SU(2)$ instanton.
A suitable $\phi$ satisfies
\begin{equation}
\left(
\partial_{z_1}^2+\partial_{z_2}^2\right)\phi(z)
=-2\pi\kappa\left[
\delta^2(z-\omega)-\delta^2(z+\omega)\right],
\end{equation}
where $\kappa$ is a constant and $\pm\omega$
are the positions of the two singularities
(we have used translational invariance to shift the `centre of gravity'
of the singularities to the origin).
The delta functions should be read as periodic (with respect to
$z_1\rightarrow z_1 +2\pi/L_1$ and $z_2\rightarrow
z_2+2\pi/L_2$).
Physically, the Nahm potential describes two Aharonov-Bohm fluxes
of strength $\kappa$ and $-\kappa$
threading the dual torus.
They must have equal and opposite strength to ensure a periodic $\hat A$.
We may assume that $\kappa$ lies between $0$ and $1$
since it is possible via a (singular) gauge transformation
to shift $\kappa$ by an integer amount
(under such a transformation the \sl total \rm flux through
$\tilde\T^2$ remains zero).

One can write $\phi$  explicitly in terms of Jacobi theta functions
\begin{equation}\label{phi}
\phi(z)=\frac{\kappa}{2}\left(
\log
\frac{\left|\theta\left(
(y+\omega_1+i\omega_2)\frac{L_1}{2\pi}
+\frac{1}{2}+\frac{iL_1}{2L_2},\frac{iL_1}{L_2}\right)\right|^2}
{\left|\theta\left(
(y-\omega_1-i\omega_2)\frac{L_1}{2\pi}
+\frac{1}{2}+\frac{iL_1}{2L_2},\frac{iL_1}{L_2}\right)\right|^2}
+\frac{iL_1 L_2 \omega_2}{\pi} (y-\bar y)-
2\omega_2 L_1
\right).
\end{equation}
The theta function is defined as
\begin{equation}
\theta(w,\tau)=
\sum_{n=-\infty}^\infty
e^{i\pi n^2\tau+2\pi in w},
~~~~~~~\hbox{Im}\,\tau>0,\end{equation}
and has the periodicity properties
$
\theta(w+1,\tau)=\theta(w,\tau)$ and
$\theta(w+\tau,\tau)=e^{-i\pi \tau-2\pi i w}
\theta(w,\tau)$.
In each cell $\theta(w,\tau)$ has a single zero
located at the centre of the torus ($w=\h+\h\tau$).
We have chosen the constant term in (\ref{phi}) so that
the integral of $\phi$ over the dual torus is zero.
This renders $\phi(z)$ an odd function, $\phi(-z)=-\phi(z)$.  
The Nahm potential derived from (\ref{phi})
 was obtained in \cite{Ford:2000zt}
via the ADHM
construction. Here the flux strength $\kappa$ is related to the
ADHM
`size', $\lambda$, of an instanton centred at $x_\mu=0$, 
\begin{equation}\label{size}
\kappa=\frac{\pi\lambda^2}{L_1L_2}.
\end{equation}
Although we do not directly
 use  the ADHM formalism in the present paper
the relation (\ref{size}) proves useful in interpreting our results.

The $SU(2)$ instanton we seek, $A_\mu(x)$, is the Nahm transform
of the abelian potential $\hat A(z)$
\begin{equation}\label{inversenahm}
A_\mu^{pq}(x)=\int_{\tilde T^2}d^2z\,
{\psi^p}^\dagger(z;x)
\frac{\partial}{\partial x^\mu}\psi^q(z;x),\label{inverse}
\end{equation}
where 
the $\psi^p(z;x), p=1,2$ are orthonormal zero modes of
\begin{equation}\label{weylop}
-\frac{i}{2} D_x^\dagger(\hat A)=
\left(
\begin{array}{cc}
\h \bar x_\perp &
\partial_y+\partial_y\phi-\frac{i}{2}\bar x_{||}
\\
\partial_{\bar y}-\partial_{\bar y}\phi-\frac{i}{2}x_{||}
&\h x_\perp
\end{array}\right). 
\end{equation}
In addition to the complex coordinates $y$, $\bar y$ on $\tilde\T^2$
we have introduced two sets of complex coordinates
for $\T^2\times\R^2$; in the `parallel' directions
$x_{||}=x_1+ix_2$, $\bar x_{||}=x_1-ix_2$,
and in the `transverse' non-compact directions
$x_\perp=x_0+ix_3$, $\bar x_\perp=x_0-ix_3$.

When $x_\perp=0$ the Weyl equation
decouples and the two zero modes have a simple form \cite{Ford:2000zt}
\begin{equation}\label{modes}
\psi^1(z;x)=\left(\begin{array}{cr}
0 \\ e^{-\phi(z)}G_+(z-\omega)
\end{array}\right),~~~~
\psi^2(z;x)=\left(\begin{array}{cr}
e^{\phi(z)}G_-(z+\omega)\\
0
\end{array}\right),
\end{equation}
where $G_\pm(z)$ are periodic Green's functions satisfying
\begin{equation}
\left(-i\partial_{y}-\h\bar x_{||}\right)G_+(z)
=\h\delta^2(z),~~~~
\left(-i\partial_{\bar y}-\h x_{||}\right)
G_-(z)=\h\delta^2(z).
\end{equation}
 $G_-(z)$ has a theta function representation
\begin{equation}\label{szeboe}
G_-(z)=\frac{iL_1}{4\pi^2}
e^{\h ix_{||}(\bar y-y)}
\frac{\theta'(\h+\frac{iL_1}{2L_2},\frac{iL_1}{L_2})\theta
(y+\h+\frac{iL_1}{2L_2}-\frac{i}{L_2}x_{||},\frac{iL_1}{L_2})}{
\theta(\h+\frac{iL_1}{2L_2}-\frac{i}{L_2}x_{||},\tau)
\theta(y+\h+\frac{iL_1}{2L_2},\frac{i L_1}{L_2})},
\end{equation}
with $\theta'(w,\tau)=\partial_w \theta(w,\tau)$.
The corresponding result for
$G_+(z)$ can be obtained via  $G_+(z)=G_-^*(-z)$.
The zero modes have square-integrable singularities at both fluxes:\\[-1ex]
\begin{center}
\begin{tabular}{r|l|l|}
\hline\hline\\[-4.3ex]
$\quad{}                         \quad{}$&
$\di \qquad\qquad{}  z\sim \omega          $&
$\di \qquad{}\qquad{}   z\sim-\omega       $\\[.5ex] \hline\\[-4.2ex] 
$\di |\psi^1|^2 $ 
&$\di \quad{}\propto|y-\omega_1-i\omega_2|^{2(\kappa-1)}$
&$\di \quad{}\propto |y+\omega_1+i\omega_2|^{-2\kappa}$\\[.5ex] 
\hline\\[-4.2ex]
$\di |\psi^2|^2 $ 
&$\di \quad{}\propto|y-\omega_1-i\omega_2|^{-2\kappa}$ 
&$\di \quad{}\propto |y+\omega_1+i\omega_2|^{2(\kappa-1)}$
 \\[.3ex]\hline\hline
\end{tabular}
\end{center}
\begin{center}
  {\small {\bf Table 1:} Singularities of the $x_\perp=0$
zero modes
}\end{center} 
There are two values of $x_{||}$ (in each copy
of $\T^2$) for which Tab.~1 does not hold.  When
$x_{||}=0$ the Green's function's $G_\pm$ do not exist.  This does not
mean there are no zero modes, indeed one can see that
\begin{equation}
\psi^1(z;x=0)=
\left(\begin{array}{cr}
0\\e^{-\phi(z)}\end{array}\right),~~~~~~~
\psi^2(z;x=0)=
\left(\begin{array}{cr}
e^{\phi(z)}\\
0
\end{array}\right),
\end{equation}
are solutions of the $x_{||}=0$ Weyl equation.
$\psi^1$ has the expected square-integrable singularity
at $z=-\omega$,
but for $z=\omega$, $\psi^1(z;x=0)$ is zero.
On the other hand $\psi^2(z,x=0)$ diverges at $z=\omega$ but not at
 $z=-\omega$.
The finiteness of $G_-(2\omega)$ was used to derive 
Tab.~1. But if it is \sl zero, \rm
$\psi^1$ will lose its singularity at $z=-\omega$.
From the theta function representation of $G_-(z)$
one can see that there is exactly one value of 
$x_{||}$ in $\T^2$
where this occurs, $\pi x_{||}=-iL_1 L_2(\omega_1+i\omega_2)$.
We wish to investigate whether these `soft'
points are
 constituent locations.

Notice that the singularity profiles of $\psi^1$ and $\psi^2$ are
exchanged under the replacement
$\kappa\rightarrow 1-\kappa$.
This suggests that the constituents are exchanged under this
mapping. That is, if there are indeed lumps at 
the two points, then $\kappa\rightarrow 1-\kappa$
swaps the two lumps.
The following result formalises this idea
\begin{equation} \label{duality}
F_{\mu\nu}(x_{||},x_\perp,\kappa)=
V^{-1}(x)
F_{\mu\nu}\left(-x_{||}+\frac{L_1 L_2}{i\pi}(\omega_1+i\omega_2),-x_\perp,
1-\kappa\right)V(x),
\end{equation}
where $V(x
)$ is some $U(2)$ gauge transformation. 
The proof of \Eq{duality} goes as follows: make the change of variables
$z\rightarrow -z$ in (\ref{inversenahm}). The zero mode $\psi^p(-z;x)$
satisfies the same Weyl equation as $\psi^p(z;x)$ except
that the signs of $\kappa$ and the $x_\mu$ are flipped.
Under periodic gauge
transformations 
$\hat A(z,-\kappa)$ and $\hat A(z,1-\kappa)$
 are equivalent up to
a constant potential 
\begin{equation}
\hat A_\mu(z,-\kappa)= U^{-1}(z)\left(
\partial_{\mu}+\hat A_\mu(z,1-\kappa)+B_\mu\right)U(z), 
\end{equation}
where $\pi B_1=-iL_1 L_2\omega_2$
and $\pi B_2=iL_1 L_2 \omega_1$.
In the Weyl equation $B_\mu$ can be absorbed into $x_1$ and
$x_2$.
Thus $U^{-1}(z)\psi^p(z,-x_{||}-i L_1 L_2(\omega_1+i\omega_2)/\pi,
-x_\perp)$ satisfies the same Weyl equation as
$\psi^p(-z;x_{||},x_\perp)$.

We have established that any lumps centred at the constituent points have
a simple exchange property.
To actually see the lumps we must compute the field strengths.
We can do this explicitly in the $x_\perp=0$ slice,
since here we have the exact zero modes (\ref{modes}).
Before we insert these modes into  (\ref{inversenahm}), 
they must be normalised.
This can be done by dividing both modes by $\sqrt{\rho}$, where
\begin{equation}\label{rho}
\rho=\int_{\tilde T^2}d^2z \,
e^{2\phi(z)}|G_-(z+\omega)|^2.
\end{equation}
In fact, one can write two components of the gauge potential
solely in terms of this normalisation factor
\begin{equation}
A_{x_{||}}=-\h \tau_3\partial_{x_{||}}\log\rho,~~~~~~~~~~
A_{\bar x_{||}}=\h \tau_3\partial_{\bar x_{||}}\log\rho,
\end{equation}
so that
\begin{equation}\label{F||}
F_{x_{||}\bar x_{||}}=\tau_3\partial_{x_{||}}\partial_{\bar x_{||}}
\log\rho.\end{equation}
The computation of the other components is more involved since the 
derivatives
of the zero modes with respect to $x_\perp$ are required.
However, the field strengths can be  more
easily accessed
via Green's function techniques.
Here we just quote the results; more details will be given elsewhere
\begin{equation}\label{F|_}
F_{x_{||}\bar x_\perp}=2\pi i(\tau_1+i\tau_2)\,
\kappa\rho\,\partial^2_{x_{||}}\frac{\nu}{\rho}.
\end{equation}
where
\begin{equation}\label{nu}
\nu=\int_{\tilde T^2} d^2z\,
G_+(\omega-z)e^{2\phi(z)}G_-(z+\omega).
\end{equation}
The other components are fixed by self-duality, i.e.
$F_{x_\perp\bar x_\perp}+F_{x_{||}\bar x_{||}}=0$ and
$F_{x_{||} x_\perp}=0.$
Note that $\rho$ is dimensionless, real and periodic
(with respect to $x_1\rightarrow x_1+L_1$, $x_2\rightarrow x_2+L_2$),
while $\nu$ is dimensionless, complex and periodic up
to constant
phases.
Both  $\rho$ and $\nu$ diverge at
$x_{||}=0$, but the field strengths should not.
A careful analysis of $\rho$ and $\nu$ in the neighbourhood of
$x_{||}=0$ shows that the field strengths are well defined at this point.
Alternatively, one can just note that $\rho$ and $\nu$ are well behaved
at the second soft point
and invoke (\ref{duality}).

\begin{figure}
\begin{picture}(100,250)(-250,0)
\put(-200,30){\epsfig{file=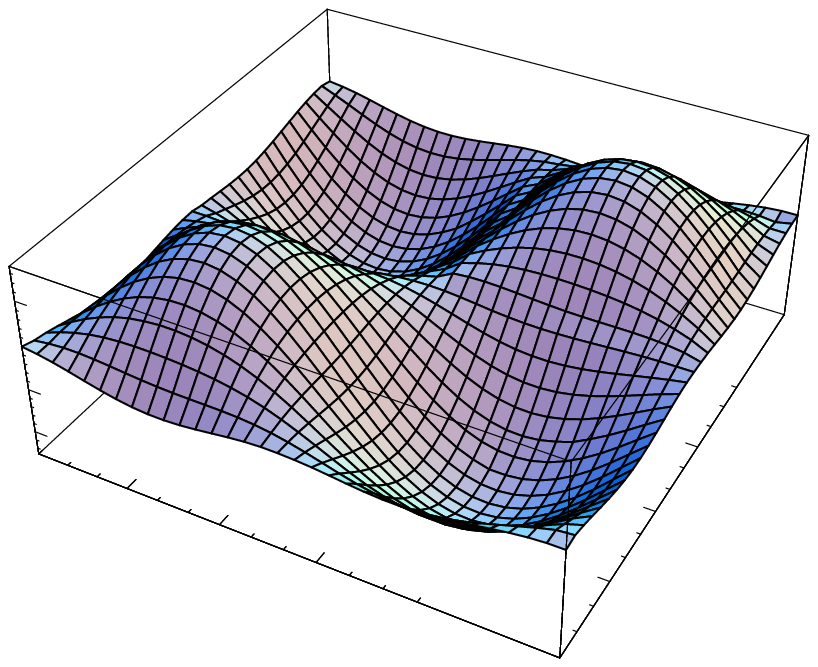,width=.7
\hsize}}
\put(-275,150){$\bf -\s012 \tr\,
 F^2$}
\put(-206,120){\small 0.1}\put(-210,138){\small 0.2}
\put(-213,157){\small 0.3}\put(-216,174){\small 0.4}
\put(-130,40){$\bf  x_2$}  
\put(-208,95){$ -\s012\pi$}\put(-150,76){ $ 0$}
\put(-98,56){$ \s012\pi$}
\put(-45,40){ $ \pi$}\put(12,16){ $ \s032\pi$}
\put(115,80){ $\bf  x_1$}
\put(30,28){ $ -\s012\pi$}\put(58,64){ $ 0$}
\put(78,100){ $ \s012\pi$}\put(102,132){ $ \pi$}\put(117,166){ $ \s032\pi$}
\end{picture}
\caption{Plot of action density for $x_\perp=0$ and 
$\kappa=\s012$  and $\omega_1=\omega_2=\s014$.
}
\label{action1}
\end{figure}

\begin{figure}
\begin{picture}(100,300)(-250,0)
\put(-200,30){\epsfig{file=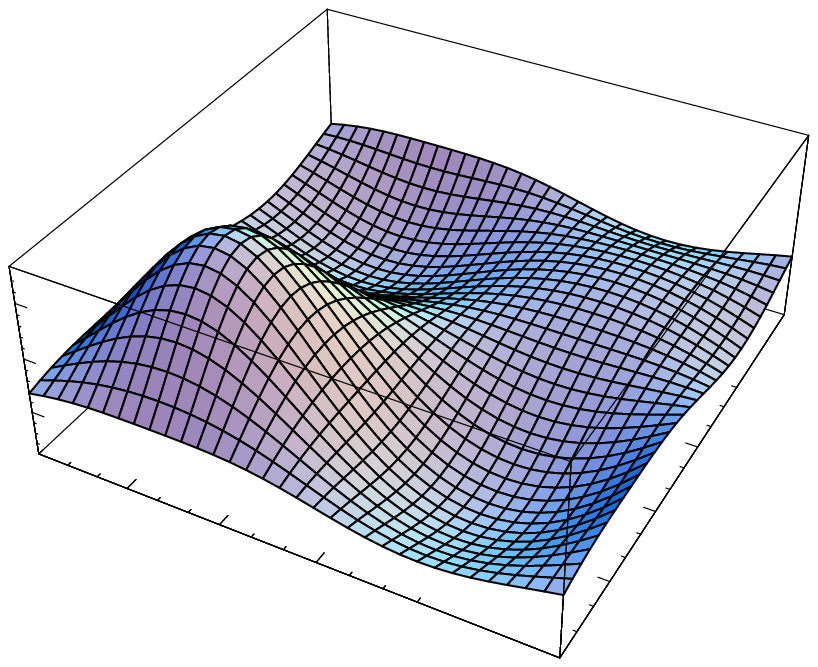,width=.7
\hsize}}
\put(-275,150){$\bf -\s012 \tr\,
 F^2$}
\put(-208,128){\small 0.2}
\put(-213,151){\small 0.4}
\put(-216,174){\small 0.6}
\put(-130,40){$\bf  x_2$}  
\put(-208,95){$ -\s012\pi$}\put(-150,76){ $ 0$}
\put(-98,56){$ \s012\pi$}
\put(-45,40){ $ \pi$}\put(12,16){ $ \s032\pi$}
\put(115,80){ $\bf  x_1$}
\put(30,28){ $ -\s012\pi$}\put(58,64){ $ 0$}
\put(78,100){ $ \s012\pi$}\put(102,132){ $ \pi$}\put(117,166){ $ \s032\pi$}
\end{picture}
\caption{Plot of action density for $x_\perp=0$ and 
$\kappa=\s0{7}{16}$  and $\omega_1=\omega_2=\s014$.
}
\label{action2}
\end{figure}

\begin{figure}
\begin{picture}(100,280)(-250,0)
\put(-200,30){\epsfig{file=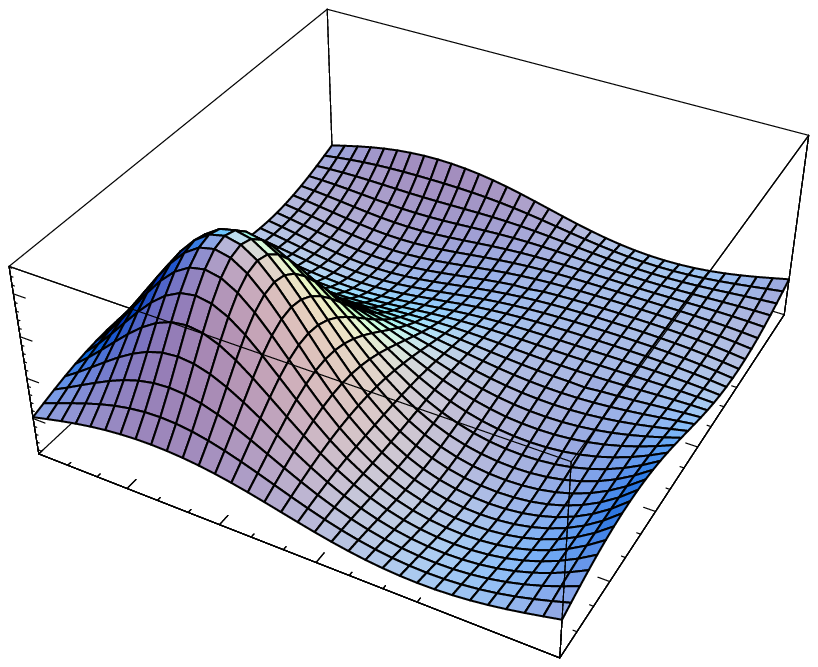,width=.7
\hsize}}
\put(-275,150){$\bf -\s012 \tr\,
 F^2$}
\put(-213,128){\small 0.25}\put(-210,144){\small 0.5}
\put(-218,159){\small 0.75}\put(-216,176){\small 1.0}
\put(-130,40){$\bf  x_2$}  
\put(-208,95){$ -\s012\pi$}\put(-150,76){ $ 0$}
\put(-98,56){$ \s012\pi$}
\put(-45,40){ $ \pi$}\put(12,16){ $ \s032\pi$}
\put(115,80){ $\bf  x_1$}
\put(30,28){ $ -\s012\pi$}\put(58,64){ $ 0$}
\put(78,100){ $ \s012\pi$}\put(102,132){ $ \pi$}\put(117,166){ $ \s032\pi$}
\end{picture}
\caption{Plot of action density 
for $x_\perp=0$ and $\kappa=\s038$, $\omega_1=\omega_2=\s014$.}
\label{action3}
\end{figure}

Since we have all field strengths within the $x_\perp=0$ slice
we can compute the action density
\begin{equation}
\label{actiondensity}
-\h \hbox{Tr} F_{\mu\nu}F^{\mu\nu}=
16\left(\partial_{x_{||}}\partial_{\bar x_{||}}\log\rho
\right)^2+
128\pi^2\kappa^2\rho^2\left|
\partial_{x_{||}}^2\frac{\nu}{\rho}\right|^2.
\end{equation}
We have provided integral representations of $\rho$ and $\nu$ where
the integrands are expressible in terms of standard functions.  Using
these results, we have made numerical plots of the action density for
various values of $\kappa$ and $\omega$, and for the periods we have
taken $L_1=L_2=2\pi$, i.e. a square torus.  
The most symmetric case,
$\kappa=\h$ and $\omega_1=\omega_2=\frac{1}{4}$ is plotted in Fig.~
\ref{action1}.  Two equal sized lumps are observed, one at
$x_{||}=0$, the other at $x_{||}=(1+i)\pi$.
Decreasing $\kappa$ at fixed $\omega$ reduces the size of the
$x_{||}=0$ constituent but \it increases \rm its contribution to the
action.
Conversely, the second constituent becomes larger but contributes less
to the action density. These two effects tend to (rather quickly)
flatten the second peak as $\kappa$ is reduced.
 Here  we provide plots with $\kappa=\frac{7}{16}$ (Fig.
\ref{action2}) and $\kappa=\frac{3}{8}$ (Fig.~\ref{action3}), again
with $\omega_1=\omega_2=\frac{1}{4}$.
Already at $\kappa=\frac{3}{8}$ we see that the $x_{||}=(1+i)\pi$
constituent is more of a plateau than  a peak.
If $\kappa$ is decreased much further the first peak will dominate
completely; this peak will be essentially
 that of a BPST instanton on $R^4$ with a scale parameter
given by the ADHM formula (\ref{size}).
Indeed, if we take  (\ref{size}) at face value, and use
(\ref{duality}),  our two constituents are BPST 
instanton cores, one with scale $\lambda
=\sqrt{\kappa L_1 L_2/\pi}$ at $x_{||}=x_\perp=0$,
the other with scale $
\lambda
=\sqrt{(1-\kappa)L_1 L_2/\pi}
$ at the second soft point.
When $L_1=L_2$ at least one $\lambda$ is of the order of the size of
the torus in line with the strong overlap seen in the plots.

The peaks are best resolved at $\kappa=\h$, when the
constituents are identical.
The overlap can be avoided by taking one period, say
$L_2$, smaller than the other, and then taking a sufficiently large 
constituent separation in the $x_1$ direction.
However, then the size of one or both cores will exceed $L_2$,
tending to suppress or smooth out
the dependence on $x_2$. In this case the constituents
become \it periodic monopoles. \rm
Actually, it is possible to see some hint of this monopole limit
even for a square torus; taking $\omega_1=\frac{1}{4}$,
$\omega_2=0$, $\kappa=\h$ with $L_1=L_2=2\pi$ the two constituents
(at $x_{||}=0$ and $x_{||}=i\pi$) overlap so strongly
that they merge into a monopole worldline, in that the dependence
of the field strengths on $x_2$
is very weak.

\begin{figure}
\centering
\begin{picture}(100,250)(100,0)
\epsfxsize=10cm\epsffile{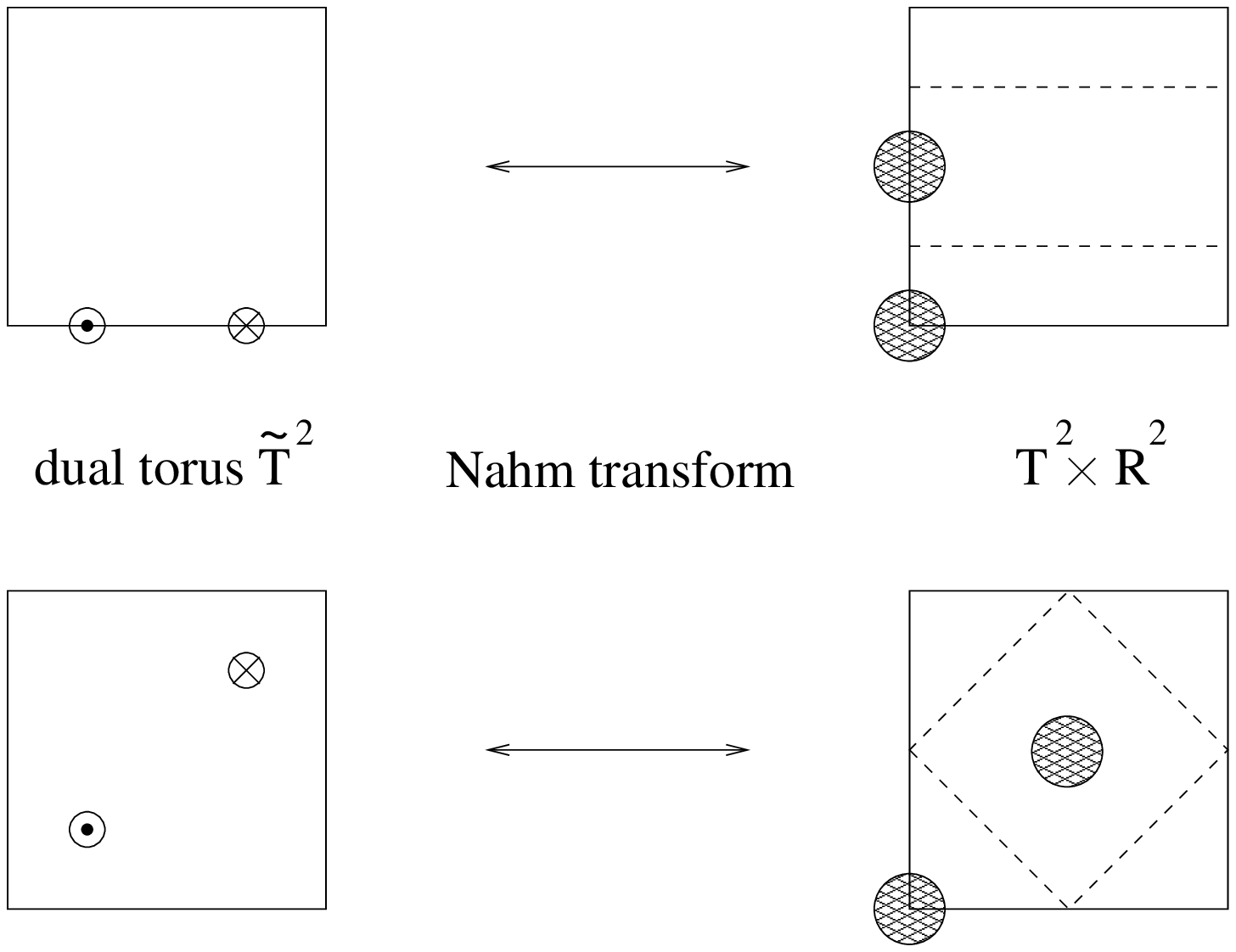}
\end{picture}
\caption{When $\kappa=\h$, the
choices $\omega_1=\h\pi/L_1$,
$\omega_2=0$ and
$\omega_1=\h\pi/L_1$,
$\omega_2=\h\pi/L_2$
yield constituent separations
allowing the torus to be cut (dashed line) to
yield charge $\h$ instantons.
}
\label{constituents}
\end{figure}

The $\kappa=\frac{1}{2}$ case has another interesting feature.  If we
choose the constituent locations so that they are separated by half
periods the charge one instanton can be `cut' to yield a charge
$\frac{1}{2}$ instanton (see also \cite{Gonzalez-Arroyo:1998ia}).
This happens when $(\omega_1,\omega_2)$ is $(\h\pi/ L_1,0)$, $(0,\h
\pi/ L_2)$ or $(\h\pi/ L_1,\h\pi/ L_2)$; the former and the latter are
shown in Fig.~ \ref{constituents}.  After cutting we have a twist
$Z_{12}=-1$ in the half torus.  But to get a half-integer topological
charge we also require a twist in the non-compact directions,
$Z_{03}=-1$, since non-orthogonal twists are a prerequisite for
fractional instanton number.  Far away from $x_\perp=0$, the potential
must be a pure gauge
\begin{equation}
A_\mu(x)\rightarrow V^{-1}(x_1,x_2,\theta)\partial_\mu V(x_1,x_2,\theta)
~~~~~~r\rightarrow \infty,
\end{equation}
where $x_\perp=re^{i\theta}$.  The non-compact twist, $Z_{03}=-1$,
translates into a double-valued gauge function,
$V(x_1,x_2,\theta+2\pi)=-V(x_1,x_2,\theta)$.  However, the explicit
results we have given are restricted to the $r=0$ slice.  What can we
say about the full $r\neq 0$ potential?  It can be argued that as for
$r=0$ (as well as $SU(2)$ calorons) two functions, one real the other
complex, are sufficient
\begin{eqnarray}
A_{x_{||}}&=&
-\frac{\tau_3}{2}\partial_{x_{||}}\log\rho
-2\pi i(\tau_1-i\tau_2)\kappa \rho \partial_{\bar x_\perp}
\frac{\nu^*}{\rho},
\nonumber\\
A_{x_{\perp}}&=&-
\frac{\tau_3}{2}\partial_{x_{\perp}}\log\rho
+2\pi i(\tau_1-i\tau_2)\kappa \rho \partial_{\bar x_{||}}
\frac{\nu^*}{\rho}.
\end{eqnarray}
Here $\rho$ and $\nu$ depend on $r$.  The $\rho$ and $\nu$
respectively given in (\ref{rho}) and (\ref{nu}) should be understood
as the zeroth order terms of power series in $r^2$ for the `full'
$\rho$ and $\nu$.  Assuming that $\nu$ falls off exponentially for
large $r$, a decay $\rho\sim r^{-1}$, would account for the twist.

Doubly periodic charge $\frac{1}{2}$ instantons have also been found
in simulations \cite{Gonzalez-Arroyo:1998ez,Montero:rb}.  Many
qualitative features reported are in accord with our findings.  In
particular, we agree that these objects are single lumps with radial
symmetry, and that the action density is never zero in the $x_\perp=0$
plane.  Another striking observation in
\cite{Gonzalez-Arroyo:1998ez,Montero:rb} is that the action density
decays exponentially for large $r$.  To make a comparison here, more
information about the
large $r$ properties of $\rho$ and $\nu$ is required. \\[.5ex]

\noindent{\bf Acknowledgements}\\
We thank F.~Bruckmann, M.~Engelhardt, O.~Jahn, T.~Tok and P. van Baal
for helpful discussions.  C.F. was supported through a European
Community Marie Curie Fellowship (contract HPMF-CT-2000-00841).\\[.5ex]

\eject 

\end{document}